# Inverse-Compton emission from halos around stars

E. ORLANDO[1], A.W. STRONG[1]
[1] *Max-Planck- Institut für extraterrestrische Physik, Postfach 1312, D-85741 Garching, Germany*
*elena.orlando@mpe.mpg.de*

**Abstract:** Inverse Compton scattering by relativistic electrons produces a major component of the diffuse emission from the Galaxy. The photon fields involved are the cosmic microwave background and the interstellar radiation field from stars and dust. Calculations of the inverse Compton distribution have usually assumed a smooth ISRF, but in fact a large part of the Galactic luminosity comes from the most luminous stars which are rare. Therefore we expect the ISRF, and hence the inverse Compton emission, to be clumpy. We also show that some of the most luminous stars may be visible to GLAST. In this paper we give an update on our previous work including examples of the intensity distribution around stars, and the predicted spectrum of Cygnus OB2.

## Estimates of Gamma Rays from luminous stars

Calculations of the inverse Compton Galactic emission have usually assumed a smooth ISRF, but in fact a large part of the Galactic luminosity comes from the most luminous stars which are rare. Therefore we expect the ISRF, and hence the inverse Compton emission, to be clumpy [1]. We have found that the contribution of the most luminous stars is non-negligible and even individual luminous stars can contribute to the clumpiness of the emission. Moreover OB associations could be detectable by GLAST. Emission by the same process from the Sun has been detected, see [2, 3].

The value of the photon energy density obtained for bright stars (Fig 1) is above the mean interstellar value of about 1 eV cm$^{-3}$ even at 10 pc distance from the star, suggesting it contributes to clumpiness in the ISRF. The photon density as function of the distance from the star and the energy is given by

$$u(r,\lambda) = 0.5\, u_{BB}(\lambda,T)\left(1 - \sqrt{1 - (R_{STAR}/r)^2}\right)$$

with $u_{BB}(\lambda, T)$ black body photon density and T the temperature of the star.
The inverse-Compton intensity along the line of sight, in the first approximation has the form:

$$I(E_\gamma,\alpha) \propto \frac{L_{STAR}}{\alpha\, d}$$

where $L_{STAR}$ is the luminosity, d the distance and α the angular distance from the star. Then

$$F(E_\gamma) \propto L_{STAR}\, \frac{\alpha}{d}$$

is the flux integrated over a volume surrounding the star.

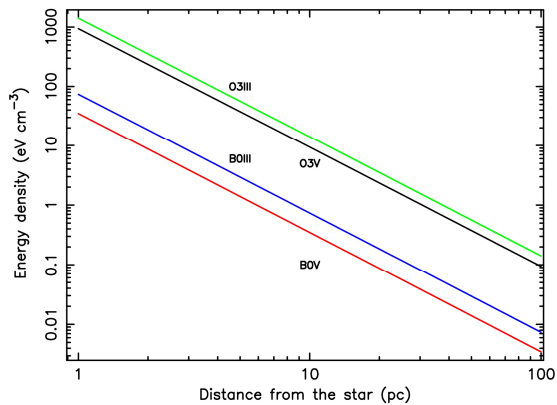

Figure 1: Photon energy density around bright stars with different spectral types. The value of the photon energy density obtained for bright stars is above the mean interstellar value of about 1 eV cm$^{-3}$ even at 10 pc distance from the star.

The inverse-Compton emission has been computed using the local interstellar electron spectrum [4] and the Klein-Nishina cross section [5].

(Explaining the diffuse Galactic emission including the 'GeV excess' requires about a factor 4 higher electron spectrum described by the "optimized" model [9]. The present estimates are therefore rather conservative).

Figure 2 shows, as an example, the cumulative flux above 100 MeV from different stellar types stars at 100 pc as a function of the angle of integration; it can be seen that the emission is extended over several degrees. The main contribution to the emission comes from more than a few pc from the star and hence is mostly beyond the influence of stellar winds [6].

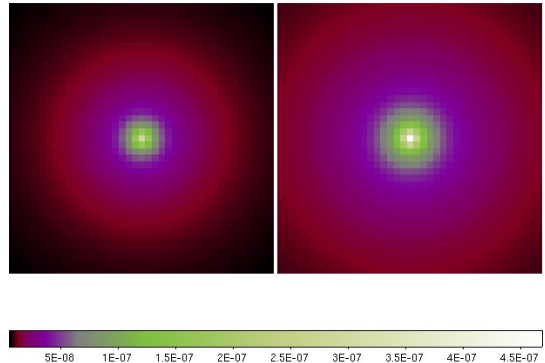

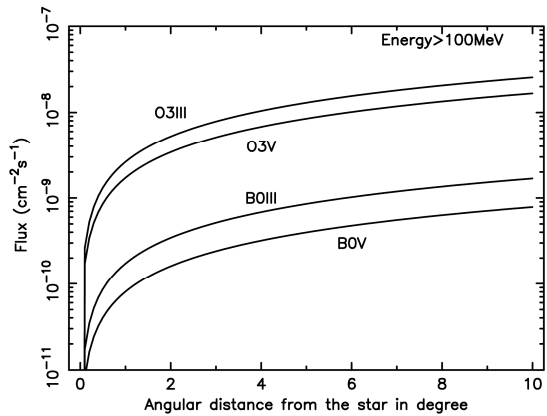

Figure 2: Cumulative flux integrated over solid angle from stars of different spectral type at 100 pc distance as a function of angle for energies above 100 MeV.

Figure 3: Expected profile of zeta Ori ($T_E$ =30500 K, Radius = 22.1 $R_{SUN}$) above 100 MeV. On the left the distance is 400 pc taken from [7], on the right the distance is 250 pc [8]. The scale shows the intensity (cm$^{-2}$ s$^{-1}$ sr$^{-1}$) above 100 MeV. The region shown is 20° wide.

## Emission from Single Stars

The extended IC emission from some nearby O stars in the Galaxy has been computed. Figure 3 shows, as an example, the intensity distribution of zeta Ori above 100 MeV with two different estimates of the distance. Along the line of sight, the minimum distance from the star is taken ~ 2 pc, to avoid the modulation of the electron spectrum due to stellar winds. This also means that the IC emission could be greater than we estimate, due to electrons accelerated in stellar winds.

In the Hipparcos Catalogue [7] the parallax of zeta Ori is 3.99 ± 0.79 mas corresponding to 250 pc distance, while in [8] the distance is 400 pc. The estimates obtained are affected by the big uncertainty of the parameters of the stars, which could mean that some might be brighter gamma ray sources than these estimates. In order to have an estimate of the relevance of the emission from single stars, we compared the estimates in Figure 3 with the EGRET data.

For comparison in the direction of zeta Ori (l = 290° and b = -30°) the value of the intensity of galactic background in the EGRET data is about 3 × 10$^{-5}$ cm$^{-2}$ s$^{-1}$ sr$^{-1}$. Since our estimates are much below the EGRET sensitivity, in [2] we investigated the detectability of bright stars for the upcoming GLAST mission.

In order to calculate the flux emitted by selected stars, the angular radius to which the flux is taken is arbitrary, but we choose a value of 5° which is a compromise between angular resolution and sensitivity of gamma-ray telescopes. The value of this angle could be further optimized for specific cases.

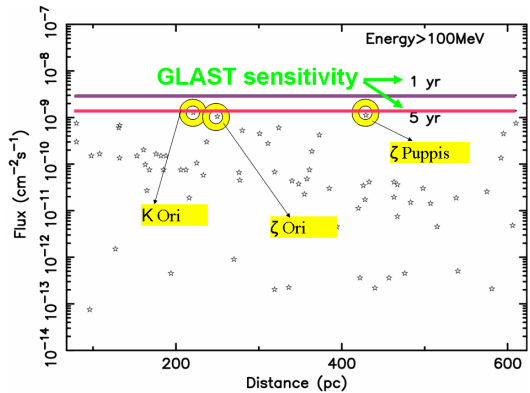

Figure 4: Gamma flux integrated over 5° from the 70 most luminous stars from the Hipparcos catalogue for E > 100 MeV, compared with the GLAST point source sensitivity for 1 and 5 years observations (horizontal lines).

The candidates were taken from the Hipparcos nearby star catalogue [10], choosing only the first 70 most luminous stars. This list includes stars up to 600 pc distance. The results are compared in Fig 4 with the GLAST point source sensitivity above 100 MeV of about $3 \times 10^{-9}$ cm$^{-2}$ s$^{-1}$ and $1.5 \times 10^{-9}$ cm$^{-2}$ s$^{-1}$ for one year and 5 years observations respectively. According to our predictions, the most gamma-ray bright stars are κ Ori (B0Iab, distance 221 pc), zeta Pup (O5Ib, distance 429 pc) and zeta Ori (O9.5Ib, distance 250 pc) with fluxes of 1.2, 1.1 and $1 \times 10^{-9}$ cm$^{-2}$ s$^{-1}$ respectively, for energy above 100 MeV and within 5°.

Another important candidate is η Carinae with $T_E$ = 30000 K, 2.3 kpc distance and luminosity of $7 \times 10^6$ solar luminosity [11]. The estimated inverse Compton flux within 5° angle is 2.2, 0.1, $0.005 \times 10^{-9}$ cm$^{-2}$ s$^{-1}$ respectively for energy >100 MeV, >1 GeV and >10 GeV.

Most of the sources are still below the GLAST threshold; however they can produce significant fluctuations on the diffuse Galactic emission.

## OB associations: Cygnus OB2

Apart from individual stars, the full stellar population will exhibit features due to clustering e.g. in OB associations. As an example, Figure 5 shows the spectrum of Cygnus OB2 centred on the cluster. 120 O stars and 2489 B stars and 6000 F stars have been distributed in a region of 2° diameter (60 pc) at 1.7 kpc distance. Data for the Cygnus association and the number of stellar components are taken from [12]. For this simulation the stars have been supposed to be O5V with $T_E$ = 42000 K and R = 12 $R_{SUN}$, B5V with $T_E$ = 15200 K and R = 3.9 $R_{SUN}$ and F5V with $T_E$ = 8180 K and R = 1.7 $R_{SUN}$. The estimated IC emission is comparable to the total IC emission from the Cygnus region and the extragalactic background [4]. The emission extends to more than a region of 1° radius, due to the size and the extension of the IC emission. The IC flux calculated from our model within 5° is 18, 1.9, $0.05 \times 10^{-9}$ cm$^{-2}$ s$^{-1}$ respectively for energy >100 MeV, >1 GeV and >10 GeV, while within 1° is 3.7, 0.3 and $0.008 \times 10^{-9}$ cm$^{-2}$ s$^{-1}$ for the previous energy ranges. This will clearly be of interest for GLAST.

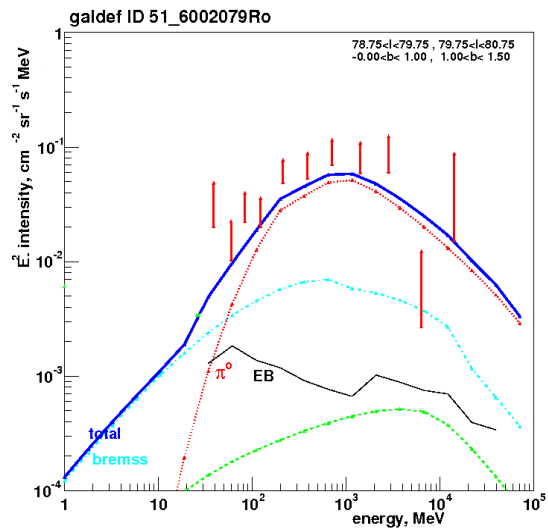

Figure 5: Gamma-ray spectrum of the Cygnus region. The green line represents the inverse-Compton emission predicted from Cyg OB2, centred on the cluster, and is comparable with the total inverse-Compton emission from the Galaxy in this region [4]. The red bars are the EGRET data, the thick black line the extragalactic background, as in [4]. The blue solid line represent the total galactic emission, the blue dashed line the bremsstrahlung and the red line the pion decay calculated with the optimized electron spectrum from GALPROP [4].

## Conclusion

We have estimated the gamma-ray emission by inverse Compton scattering of cosmic-ray electrons with the radiation field around stars.

We find that the contribution of the most luminous stars is non-negligible and OB associations and even individual luminous stars can contribute to the clumpiness of the emission, and could be detectable by GLAST. A model of the Cygnus OB2 association has been computed, showing that the estimated IC emission is comparable to the total diffuse IC emission from the Cygnus region and the extragalactic background.